\documentclass[aps,pre,preprint,onecolumn,superscriptaddress]{revtex4-1}

\usepackage{graphicx} 
\usepackage{subcaption}
\usepackage{dblfloatfix} 
\usepackage{xcolor}
\usepackage{amsmath}
\usepackage{amssymb,bm,exscale}
\usepackage{parskip}
\begin{document}
\setlength\parskip{2ex}
This manuscript is a pre-print and has not undergone peer review. We welcome any feedback.
\newpage
%\preprint{}
\title{Sensitivity of Trapping Efficiency and Relative Permeability to Experimental Methodology in Laboratory Core Flooding}

\author{Catherine Spurin}
\email{cspurin@stanford.edu}
\affiliation{Department of Energy Science \& Engineering, Stanford University, USA}
\author{Catherine Callas}
\affiliation{Department of Energy Science \& Engineering, Stanford University, USA}
\author{Takeshi Kurotori}
\affiliation{Department of Chemical Engineering, Imperial College London, UK}
\author{Hamdi A. Tchelepi}
\affiliation{Department of Energy Science \& Engineering, Stanford University, USA}
\author{Sally M. Benson}
\affiliation{Department of Energy Science \& Engineering, Stanford University, USA}

%\affiliation{Department of Earth Sciences and Engineering,
%Imperial College London, London, UK}
%\affiliation{Department of Chemical
%Engineering, Imperial College London, London, UK}
%\affiliation{Shell Global Solutions International B.V., Amsterdam,
%Netherlands}

%\thanks{}


\begin{abstract}
Understanding the migration and trapping of CO$_2$ in the subsurface is vital to geologic carbon storage projects. Traditional characterization methods employ steady-state co-injection experiments to determine relative permeability and trapping efficiency. Although laboratory studies aim to replicate reservoir conditions, co-injection experiments are often selected because they facilitate steady-state flow and reduce capillary end effects. The fundamental influence of this experimental design choice on measured petrophysical parameters remains inadequately characterized. This study presents a comparative analysis between co-injection and single-phase injection experiments, specifically investigating how experimental methodology influences both trapping efficiency and pressure differential across the core sample (which is used to calculate relative permeability). Our results demonstrate significant variations in trapping behavior between these injection techniques, suggesting that the traditional focus on co-injection experiments may overlook important physical flow phenomena. Differences between injection techniques could be strategically exploited in field applications to enhance residual trapping capacity in subsurface CO$_2$ storage operations. This work highlights the importance of understanding experimental artifacts in core flooding studies and their potential applications for improving carbon storage efficiency.

\end{abstract}

%\keywords{}

\maketitle


\section{Introduction}

The movement of multiple fluids through complex porous media is ubiquitous in nature and beyond. For example, the storage of CO$_2$ in porous subsurface rocks (geologic carbon sequestration) has been identified as a critical greenhouse gas mitigation technique \cite{ipcc2014, bui2018carbon, benson2012carbon}. Understanding the migration and trapping of CO$_2$ in subsurface reservoirs is vital to ensuring the safe, long-term storage of anthropogenic emissions \cite{krevor2015capillary, krevor2011capillary, Spurin2025commentary}.

Reservoir models are used to predict the propagation and trapping of CO$_2$ underground \cite{ringrose2013salah, bentham2005co2, ampomah2016evaluation}. Key inputs for these models are relative permeability, capillary pressure, and trapping efficiency (the relationship between initial and residual CO$_2$ saturation) \cite{kumar2005reservoir, ajayi2019review}. These properties are measured on core/centimeter-scale samples, and assumed to be representative of larger scale systems \cite{Spurin2025commentary, perrin2010experimental}. In fact, experiments are often conducted at high temperatures and pressures to be representative of reservoir conditions \cite{krevor2016impact, selem2021pore, singh2018partial, krevor2012relative, kim2018two}. These experiments are time consuming and expensive. While reservoir conditions are created in the core, typically steady-state experiments, with the co-injection of both phases, are conducted because saturation profiles are constant in time and capillary end effects are less prominent \cite{pini2017capillary, bertels2001measurement, avraam1995flow, berg2024simultaneous}. Capillary end effects are defined as a strong saturation gradient caused by the non-wetting phase preferentially leaving the core and occupying the metallic end piece. Other methods to discount strong saturation gradients have included removing the end section from analysis \cite{pini2017capillary,perrin2010experimental}, correcting for end effects prior to analysis \cite{archer1973use, jackson2018characterizing, krause2015accurate, reynolds2018multiphase}, and adding a porous plate or other rock core in sequence to the rock of interest \cite{honarpour1988relative, avraam1995flow, persoff1995two}.

Steady-state co-injection experiments have been shown to have different underlying pore scale dynamics than unsteady-state co-injection experiments, with the unsteady-state dynamics leading to more mobile CO$_2$ \cite{spurin2020real}. However, it is typical to do steady-state co-injection experiments, with resulting values applied uniformly across the entire plume, regardless of the differing dynamics that may be present. It is also assumed that the residual trapping measured in these experiments can be used to parameterize models for hysteresis in relative permeability and capillary pressure constitutive functions \cite{krevor2015capillary}. The difference between co-injection experiments, where both fluids are injected simultaneously, and single phase injection experiments, where a single fluid is injected, must be understood. While excluding the core's end section from analysis can eliminate the saturation drop caused by capillary end effects, this approach fails to address how these effects influence pressure data used in relative permeability calculations, potentially resulting in inaccurate saturation-pressure relationships. Corrections for capillary end effects have been successfully applied to homogeneous sandstone samples, but they have failed to correct end effects in heterogeneous sandstone and carbonate samples \cite{manoorkar2021observations, wenck2021simulating}. Thus, the link between heterogeneities and capillary discontinuities is still missing. A porous plate, or additional rock core can reduce end effects. But, as with cropping, it does not untangle the role of end effects in the pressure data, instead the properties of the other samples or the porous plate are now included.

If the constraints of an experiment or the difference between different injection techniques is unknown, then utilizing the data at the field scale will lead to large uncertainties in plume migration and trapping. With single phase injection more likely to be employed in practical CO$_2$  storage operations, it is critical to quantify the impact of injection methods on fluid displacement dynamics and residual trapping behaviour. If significant time and effort are being invested into conducting core-scale experiments to understand flow dynamics at the field scale, it must be certain that the experiments are representative of what is occurring at the larger scale. In this work, the difference between co-injection and single phase injection experiments is presented. Experiments were conducted using two distinct fluid pairings—gas/water and oil/water—enabling an investigation of the differences between the conventional framework (originally developed for the oil industry) and the adaptations necessary for geologic carbon sequestration applications.

\section{Materials and Methods}
\subsection{Rocks} 
\begin{figure}[h]
	\begin{center}
		\centering
		\includegraphics*[angle=0, width=0.8\linewidth]{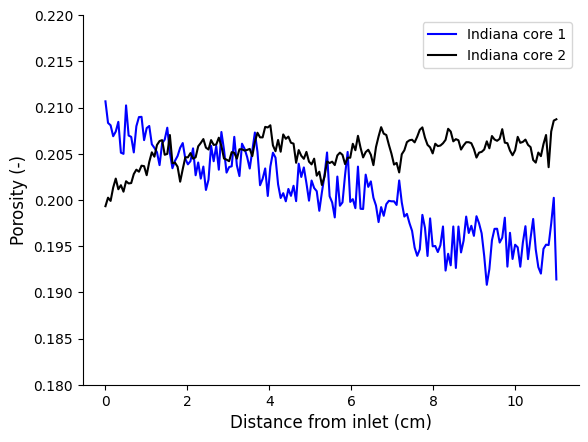}
	\end{center}
	\caption{Porosity profile for the Indiana samples used in this work. Indiana core 1 was used for the gas experiments, and the oil co-injection experiment. Indiana core 2 was used for the oil single phase injection experiment.}
	\label{porosity}
\end{figure}

Cylindrical Indiana limestone samples, 5 cm in diameter and 12 cm in length were used in this work. The same sample was used in all gas/water experiments, with the porosity along the core increasing slightly with distance along the core, as shown in Figure \ref{porosity} (Indiana core 1). For the oil/water experiments, the same Indiana core was used for the co-injection experiment. A different Indiana core (Figure \ref{porosity} Indiana core 2) was used for the single phase injection experiment. This was because removing oil from the system is time-consuming and difficult, and any remaining oil would affect the trapping efficiency reported in this work. 
Indiana is a heterogeneous limestone with a bimodal porosity distribution; the heterogeneities manifest at the millimeter scale, which is below the resolution of the images acquired in this work \cite{manoorkar2021observations}.

\subsection{Fluids, Interfacial Tension and Viscosity}
Two fluid pairings were used in this work: gas/water, and oil /water. The gas phase was nitrogen and the oil phase was decane. The wetting phase was deionized water for all experiments.
Nitrogen was used as a proxy for CO$_2$ because it has a similar viscosity and density (so the residual trapping is expected to be the same \cite{wildenschild2011exploring}), but a much lower miscibility with water. Using nitrogen allowed us to explore a gas/water system (analogous to CO$_2$ injection into a saline aquifer) while minimizing dissolution. The properties of the chosen fluids are listed Table \ref{fluid_properties}.

\begingroup
\setlength{\tabcolsep}{10pt} % Default value: 6pt
\begin{table}[]
\caption{The properties of the non-wetting phases (nitrogen or decane) and the wetting phase (water) used in this work.}
\begin{tabular}{@{}ccc@{}}
\hline
\multicolumn{1}{c}{Fluid} & \multicolumn{1}{c}{\begin{tabular}[c]{@{}c@{}}Viscosity, $\mu$\\ (Pa.s)\end{tabular}} & \multicolumn{1}{c}{\begin{tabular}[c]{@{}c@{}}Interfacial tension \\ with water, $\sigma$\ (N/m)\end{tabular}} \\ \hline
Nitrogen                  & $2.08 \times 10^{-5}$                                                                & $64 \times 10^{-3}$                                                                                                             \\
Decane                    & $83.8 \times 10^{-5}$                                                                & $47 \times 10^{-3}$                                                                                                           \\
Water                     & $95.0 \times 10^{-5}$                                                                & N/A                                                                                                                           \\ \hline
\end{tabular}
\label{fluid_properties}
\end{table}
\endgroup

\subsection{Experimental Procedure and Image Processing} 
Two injection methods were explored in this work: single phase injection and co-injection. Prior to any injection, the sample was loaded into a core holder that allows pressurization of the fluids, and placed in the medical CT scanner used to image the core and the fluid distributions within (see \citet{perrin2010experimental} for more details on the experimental apparatus). All experiments were imaged using clinical X-ray Computed Tomography (CT). This scanner creates 2D images at each location along the length of the sample. The original voxel size was 0.1875~mm $\times$ 0.1875~mm in the x and y directions, and 0.625~mm in the z direction (the direction of flow). To reduce the error in voxel-level values \cite{ni2019predicting}, the voxels were coarsen by averaging over 4 voxels in each direction. Scanning in these experiments was performed at 140 kV and 200 mA.

A confining pressure was applied that was always 2~MPa above the pressure in the core. Then 2 consecutive scans were taken to capture the rock with only air occupying the pore space. These scans were averaged to improve image quality \cite{pini2016moving, kurotori2023mixed, romano2020subcore}. This gives $CT_{ar}$, which is the CT number of the core filled with air. Then the core was then saturated with deionized water and pressurized to 8~MPa. Another 2 scans were acquired, and averaged to get $CT_{wr}$, which is the CT number of the core filled with water. From these images, the porosity ($\phi$) across the core was calculated using:

\begin{equation}
\phi = \frac{CT_{wr} - CT_{ar}}{CT_{water} - CT_{air}}
\end{equation}

where $CT_{water}$ = 0 is the CT number of the pure water phase in Hounsfield units (HU), and $CT_{air}$ = -1000~HU is the CT number of the pure air phase at atmospheric pressure and room temperature \cite{akin2003computed}. 

For the single phase injection experiments, a drainage sequence was performed first, with the non-wetting phase (the gas or oil) injected at a constant flow rate until 300 ml of non-wetting phase had been injected. The flow rates, given in Table \ref{flow_rates}, were chosen to ensure flow was in the capillary dominated regime. Then an imbibition sequence was performed, with water injected at a constant flow rate until 300 ml of water had been injected. Full CT scans were acquired during flow, and at the end of both the drainage and imbibition. 

For the co-injection experiments, a drainage sequence was performed first with both fluids injected at a constant flow rate, and fractional flow, until 300 ml of the non-wetting fluid had been injected. Then an imbibition sequence was performed, with both phases injected again, but the water constituting the majority of the total flow rate, until 300 ml of water had been injected, to allow comparable trapping relationships between the experiments. The flow rates chosen are given in Table \ref{flow_rates}. Full CT scans were acquired during flow, and at the end of both the drainage and imbibition.

Between the gas experiments the core was depressurized and flushed with water before the system was re-pressurized. This removes gas from the system, as confirmed by the wet scans prior to an experiment. For the oil co-injection experiment, the same sample was also used. For the oil single phase injection experiment, a different rock core was used. 

The saturation of the non-wetting phase ($S_{nwp}$) was calculated at the voxel level using: 

\begin{equation}
    S_{nwp} = \frac{CT_{flow} - CT_{wr}}{\phi (CT_{air} - CT_{water})}
\end{equation}

where $CT_{flow}$ is the CT number of the core filled with both fluid phases. Further details on the analysis of medical CT images is given in \citet{akin2003computed, krevor2012relative, pini2016moving, kurotori2019measuring}. For scans acquired at the end of drainage and imbibition, 2 scans were averaged to improve the signal-to-noise ratio. For scans acquired during flow, no averaging was done in the image analysis. 

\begingroup
\setlength{\tabcolsep}{10pt} % Default value: 6pt
\begin{table}[]
\caption{The flow rates chosen for the different experiments, with the injected volume of the non-wetting phase (NWP) kept constant for the drainage experiments at 300ml (or approximately 5 pore volumes). The capillary number for the experiments is also given, calculated using the methodology from \cite{Spurin2019phase}.}
\begin{tabular}{@{}ccccc@{}}
\hline
{Experiment} & \multicolumn{1}{c}{\begin{tabular}[c]{@{}c@{}}NWP flow rate\\ (ml/min)\end{tabular}} & \multicolumn{1}{c}{\begin{tabular}[c]{@{}c@{}}Water flow rate\\ (ml/min)\end{tabular}} & \multicolumn{1}{c}{\begin{tabular}[c]{@{}c@{}}Duration\\ (mins)\end{tabular}} & \multicolumn{1}{c}{\begin{tabular}[c]{@{}c@{}}Capillary \\ number\end{tabular}} \\ \hline
Gas/ water single phase drainage          & 5                                                                                              & 0                                                                                                & 60                                                                                      & $1.4 \times 10^{-8}$                                                                      \\ \hline
Gas/ water single phase imbibition        & 0                                                                                              & 5                                                                                                & 60                                                                                      & $6.0 \times 10^{-7}$                                                                      \\ \hline
Gas/ water co-injection drainage          & 5                                                                                              & 1                                                                                                & 60                                                                                      & $2.0 \times 10^{-8}$                                                                      \\ \hline
Gas/ water co-injection imbibition        & 1                                                                                              & 5                                                                                                & 60                                                                                      & $8.9 \times 10^{-8}$                                                                      \\ \hline
Oil/ water single phase drainage          & 3                                                                                              & 0                                                                                                & 100                                                                                     & $4.5 \times 10^{-7}$                                                                      \\ \hline
Oil/ water single phase imbibition        & 0                                                                                              & 3                                                                                                & 100                                                                                     & $4.9 \times 10^{-7}$                                                                      \\ \hline
Oil/water co-injection drainage           & 3                                                                                              & 0.6                                                                                              & 100                                                                                     & $5.5 \times 10^{-7}$                                                                      \\ \hline
Oil/ water co-injection imbibition        & 0.6                                                                                            & 3                                                                                                & 100                                                                                     & $5.8 \times 10^{-7}$                                                                      \\ \hline
\end{tabular}
\label{flow_rates}
\end{table}
\endgroup

\newpage

\section{Results and Discussion}

\subsection{Saturation along the core after drainage and imbibition}

\begin{figure}[!htb]
    \centering
    \begin{subfigure}[b]{0.95\linewidth}
        \centering
        \includegraphics[width=\linewidth]{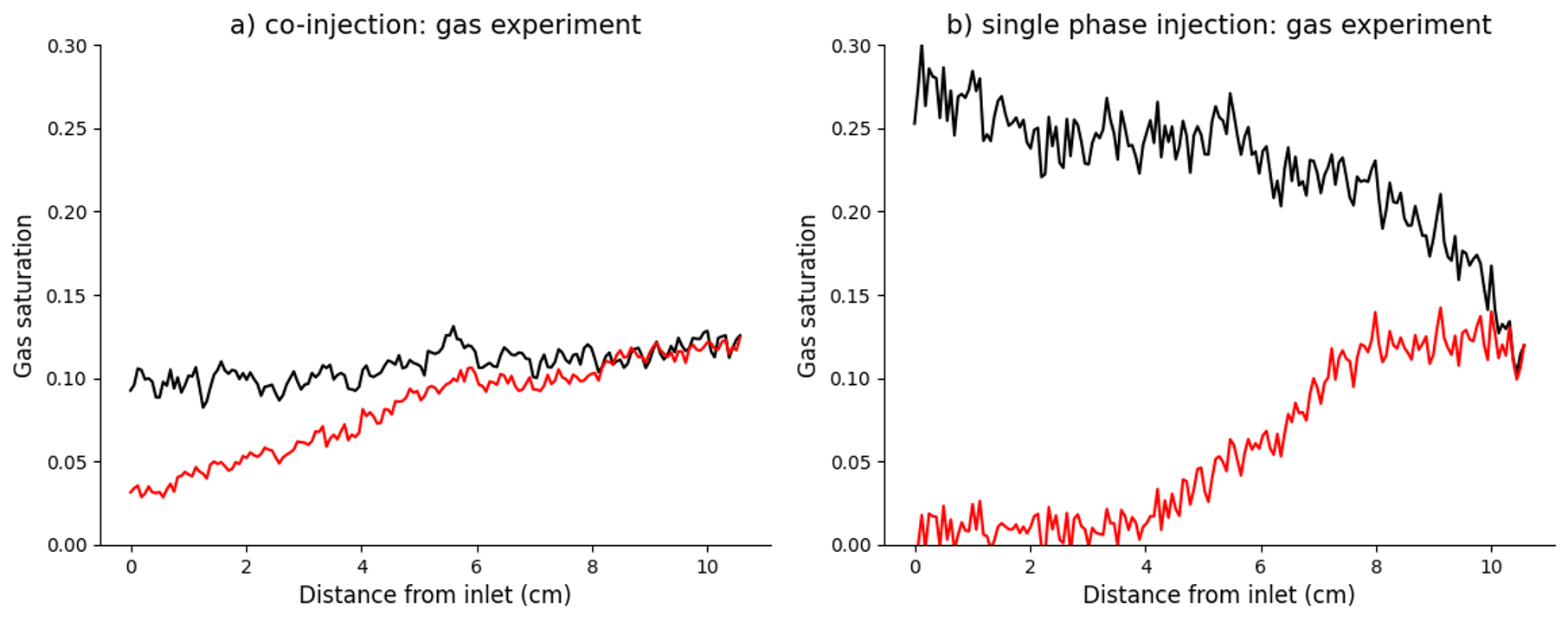}
        \label{sat_n2}
    \end{subfigure}
    
    \vspace{0.5cm} % Space between the subfigures

    \begin{subfigure}[b]{0.95\linewidth}
        \centering
        \includegraphics[width=\linewidth]{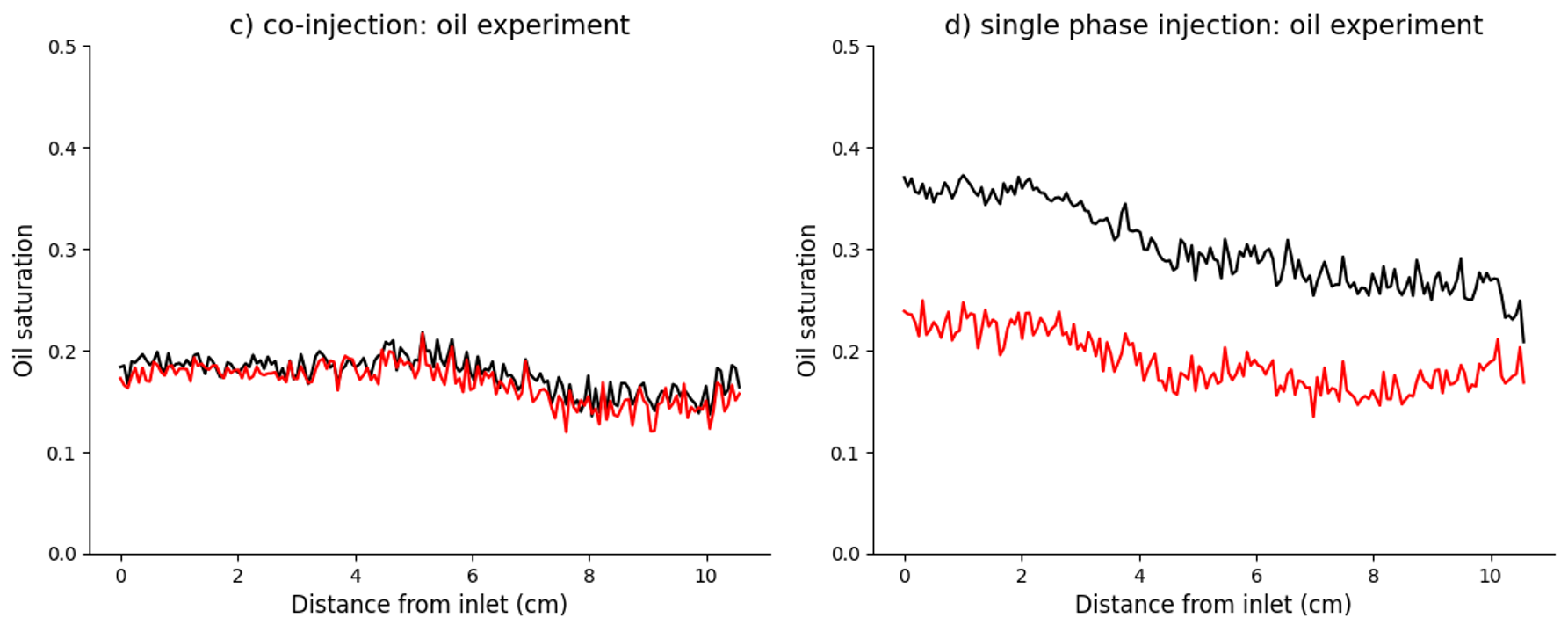}
        \label{sat_oil}
    \end{subfigure}
    
    \caption{Saturation along the core after drainage (black) and imbibition (red) for (a) co-injection of gas and water, (b) single phase injection of gas or water, (c) co-injection of oil and water, and (d) single phase injection of oil or water.}
    \label{fig:combined_saturation}
\end{figure}

Figure \ref{fig:combined_saturation} a-b) shows the saturation along the core at the end of drainage and at the end of imbibition for the gas experiments. The co-injection experiment is plotted in a), and the single phase injection experiment is plotted in b). Saturation after drainage was significantly higher for the single phase injection experiment, whereas, saturation after imbibition was similar for both experiments. This means that the difference between drainage and imbibition was larger for the single phase injection experiment. The saturation profiles suggest key differences between the injection scenarios that will be reflected in the trapping efficiency. 

Figure \ref{fig:combined_saturation} c-d) shows the saturation along the core at the end of drainage and at the end of imbibition for the oil experiments. The co-injection experiment is plotted in c), and the single phase injection experiment is plotted in d). Saturation after drainage was, again, higher for the single phase injection experiment. The difference was less extreme for the oil experiments when compared to the gas experiments. The saturation at the end of imbibition was similar for both experiments (same as the gas experiments). This means that the difference between drainage and imbibition was larger for the single phase injection experiment. The saturation profiles suggest key differences between the injection scenarios that will be reflected in the trapping efficiency but the difference will be less extreme for the oil experiments than the gas experiments.

At the end of drainage, there was a saturation gradient across the core for the single phase injection experiments (both gas and oil). This was likely due to capillary end effects, which were less extreme for the co-injection experiments. At the end of imbibition for both gas experiments, there was a strong saturation gradient across the core, increasing towards the outlet. The gas saturation was lower at the inlet, which was more extreme for the single phase injection experiment in Figure \ref{fig:combined_saturation} b. This was likely due to dissolution (although nitrogen was used because of its low miscibility in water), which would be more extreme for single phase imbibition where there was no replenishment of gas. In Figure \ref{fig:combined_saturation} b) the gas saturation was close to zero until 4~cm into the core i.e. approximately a third of the core. The drop to zero saturation at the end of the core heavily influenced the pressure drop across the core, which is discussed in Section \ref{pressure}.

\subsection{Trapping efficiency}
\begin{figure}[!h]
	\begin{center}
		\centering
		\includegraphics*[angle=0, width=1\linewidth]{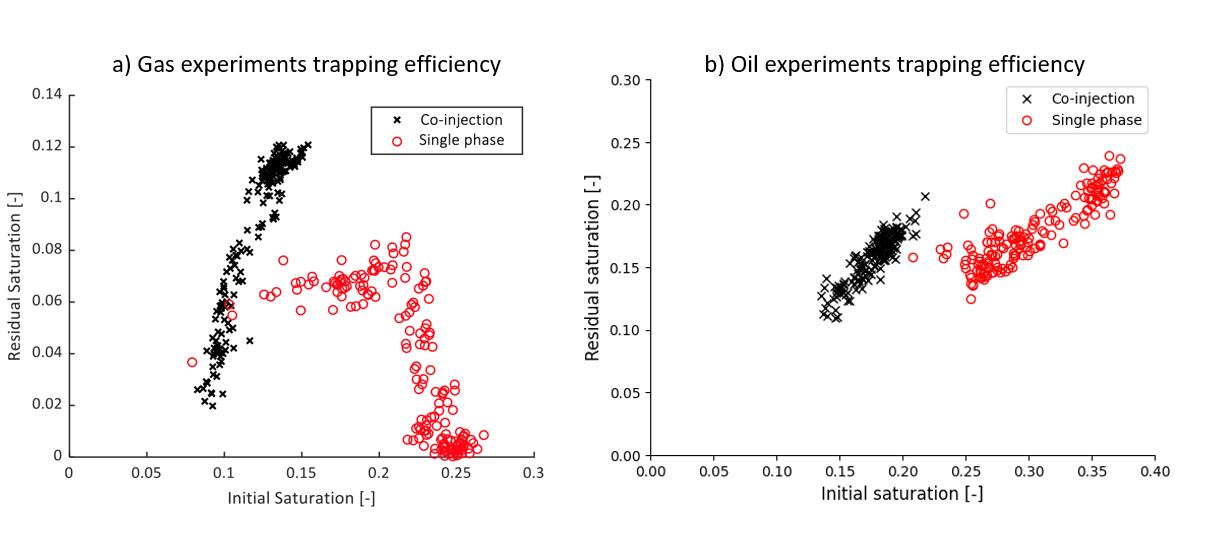}
	\end{center}
	\caption{The relationship between between initial (post drainage) and residual (post imbibition) non-wetting phase saturation for co-injection (black crosses) and single phase injection (red circles): a) gas experiments and b) oil experiments.}
	\label{trapping_curve}
\end{figure}

The trapping efficiency plots for the experiments (shown in Figure \ref{trapping_curve}) highlight the fundamental difference between the different injection techniques, and the potential errors that can arise from not considering the impact of the chosen injection technique on residual trapping. The results from the oil experiments and the gas co-injection experiment agrees with previous experimental observations by \citet{niu2015impact}, where the the slice-by-slice analysis can be fitted with a single Land trapping model. The authors argued that the end effects could be exploited as they increased the saturation range for the trapping curve, making the interpretation of the data with a Land trapping model more reliable. However, for the single phase injection gas experiment, the strong saturation gradient at the inlet creates a distinct region that cannot be modeled with a single Land trapping curve. 

It is important to note that if only the middle section of the core was analysed, or a correction was applied to the saturation to remove the perceived impact of boundary effects, the trapping curves are still dependent on the injection technique. The gas co-injection experiment has significantly higher trapping efficiency (although the initial gas saturation was lower). If co-injection experiments are conducted to estimate single phase injection, we are potentially heavily overestimating the amount of residual trapping in the subsurface, and underestimating the pore volume utilization. For the oil experiments, the difference in trapping efficiency for the different injection techniques is less extreme. However, the relationship is still shifted. The comparison is less direct in this case, as different samples were used for the different experiments.

\subsection{Pressure and saturation evolution during drainage and imbibition}
\label{pressure}

\begin{figure}[!htb]
    \centering
    \begin{subfigure}[b]{0.95\linewidth}
        \centering
        \includegraphics[width=\linewidth]{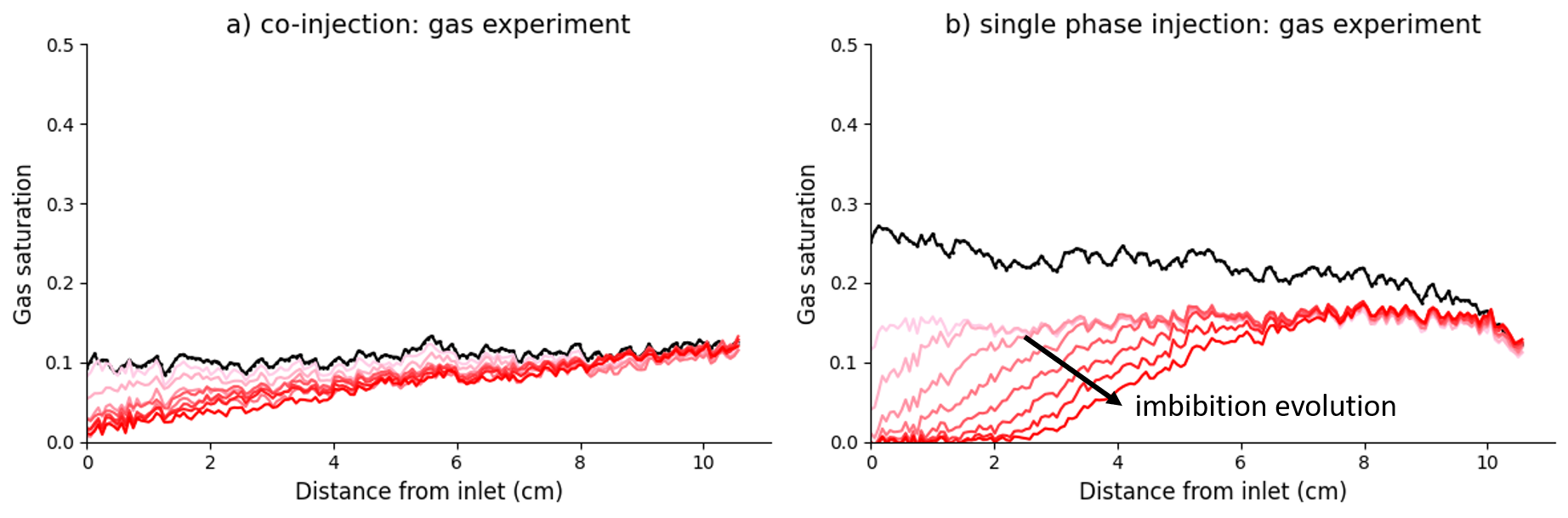}
        \label{sat_n2_evolution}
    \end{subfigure}
    
    \vspace{0.5cm} % Space between the subfigures

    \begin{subfigure}[b]{0.95\linewidth}
        \centering
        \includegraphics[width=\linewidth]{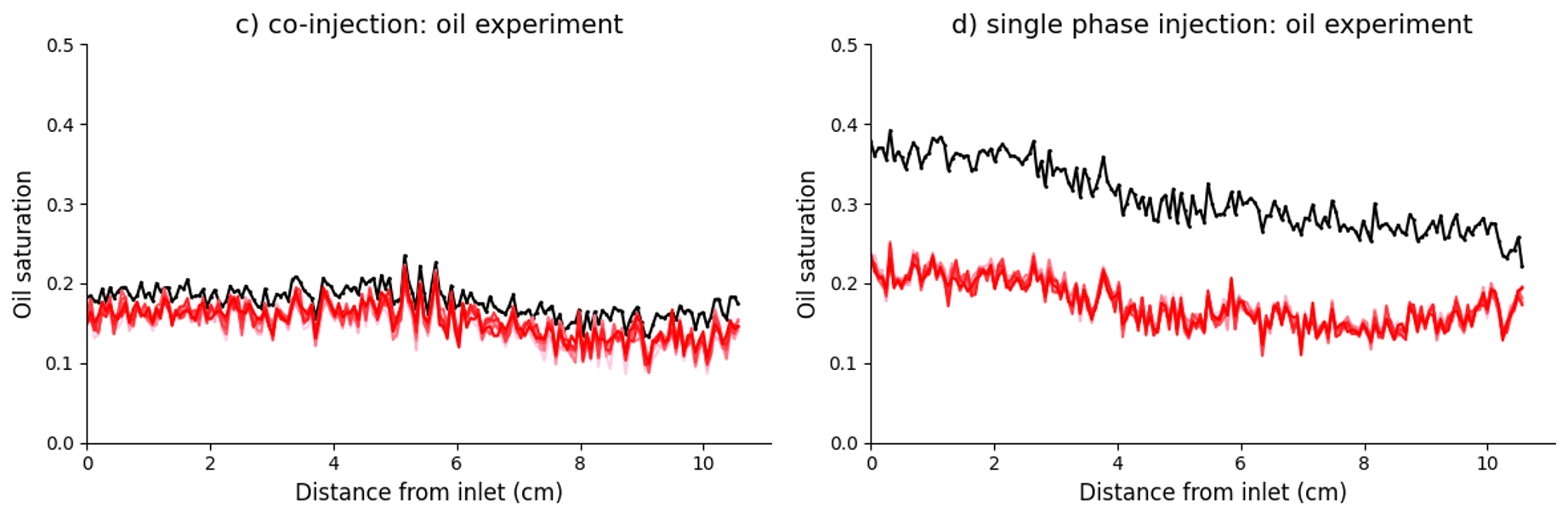}
        \label{sat_oil_evolution}
    \end{subfigure}
    
    \caption{The saturation evolution during imbibition for a) the co-injection gas experiment, b) the single phase injection gas experiment, c) the co-injection oil experiment, and d) the single phase injection oil experiment. The black line denotes the saturation at the end of drainage. The saturation during imbibition is colour coded depending on the time the scan was acquired; a lighter colour denotes an earlier scan time i.e. earlier in the imbibition experiment.}
    \label{fig:sat_evolution}
\end{figure}

The saturation values during flow are less accurate than those acquired with flow stopped. This is due to fluid movement during the acquisition of an image (which takes 20 seconds). However, the information is useful for assessing the evolution in time (which we do for imbibition as the evolution timescales are much longer than for drainage). This is shown in Figure \ref{fig:sat_evolution} for a) the co-injection gas experiment, b) the single phase injection gas experiment, c) the co-injection oil experiment, and d) the single phase injection oil experiment. The black line denotes the non-wetting phase saturation at the end of drainage. The non-wetting phase saturation during imbibition is colour coded depending on the time the scan was acquired; a lighter colour denotes an earlier scan time i.e. earlier in the imbibition experiment.  

As can be observed in Figure \ref{fig:sat_evolution}, oil saturation during imbibition was established rapidly (and so no evolution is observable). Thus, for the remainder of this section, only the gas experiments are discussed. 

\begin{figure}[h]
	\begin{center}
		\centering
		\includegraphics*[angle=0, width=0.8\linewidth]{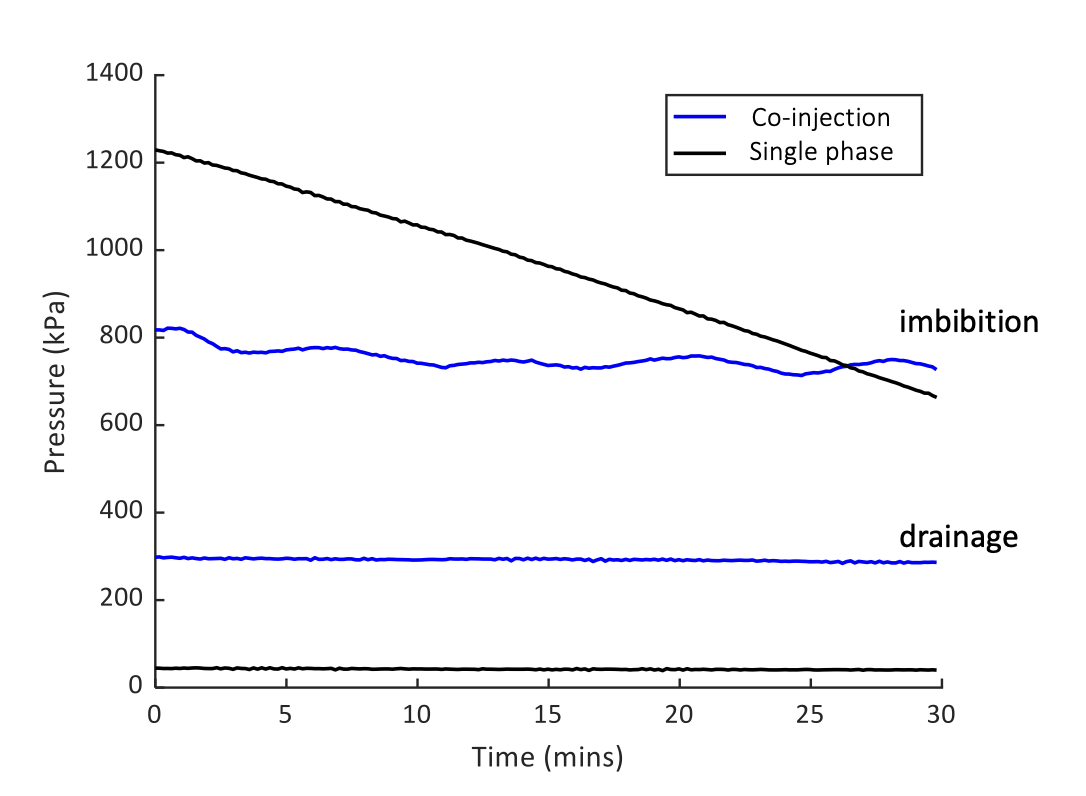}
	\end{center}
	\caption{The pressure drop across the core for the final 30 mins of injection. The blue lines show the pressure during imbibition and drainage for the co-injection experiment. The black lines show the pressure during imbibition and drainage for the single phase injection experiment. The pressure drop is higher during imbibition than drainage because trapped gas clusters increase resistance to flow.}
	\label{pressure_data}
\end{figure}

The pressure drop across the core for the final 30 minutes of the gas experiments is shown in Figure \ref{pressure_data}. The differential pressure during drainage heavily depended on the injection technique. For co-injection drainage, the gas saturation was lower, and the differential pressure was higher. This implies more tortuous pathways for the gas, causing higher resistance to flow and subsequently higher pressure drop across the core. A greater proportion of the gas remains after imbibition for the co-injection experiment, suggesting that the more tortuous gas pathways encourage more residual trapping.

For the single phase injection gas experiment, the gas saturation close to the outlet was fairly constant during the experiment (Figure \ref{fig:sat_evolution} b). This suggests that the gas saturation dropping to zero at the inlet dominated the differential pressure measurement in Figure \ref{pressure_data}. This implies that the pressure measurements, and associated relative permeability values, are heavily dominated by strong saturation gradients, providing misrepresentative values for reservoir models. Even with end effects present during co-injection imbibition, the differential pressure plateaus. This suggests that end effects do not prevent the system reaching an equilibrium, and can manifest in the pressure data in subtle ways. These different pressure responses during imbibition suggest significantly different dynamics dependent on injection technique. 

In Figure \ref{fig:sat_evolution} b), the residual gas saturation was reached quickly near the outlet. Only the region near the inlet evolves slowly. This suggests that the region influenced by end effects can be quantified by assessing the change in saturation with time. The gas saturation stabilised quicker during imbibition  for the co-injection experiment, as shown in Figure \ref{fig:sat_evolution} by more overlap at later times.

\begin{figure}[h]
	\begin{center}
		\centering
		\includegraphics*[angle=0, width=1.0\linewidth]{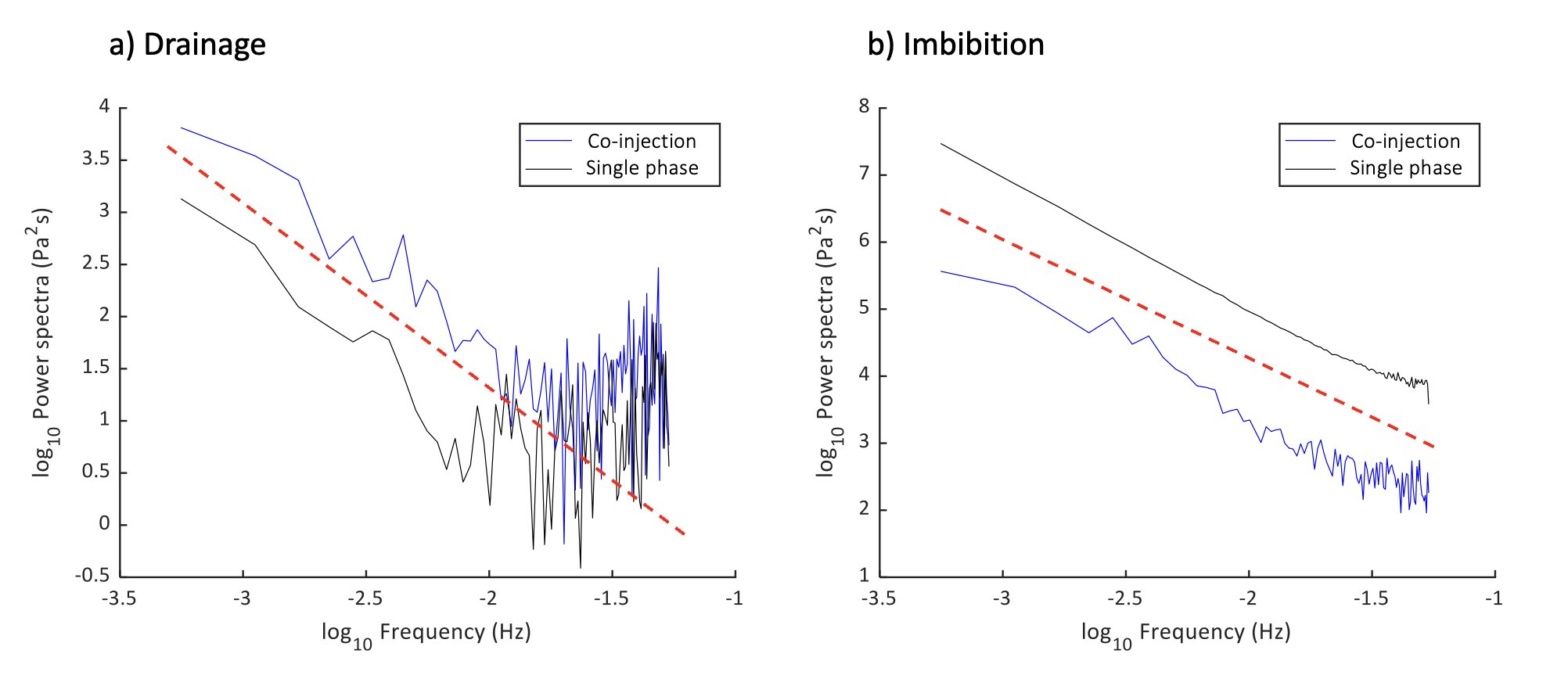}
	\end{center}
	\caption{The pressure data transformed into the frequency domain for a) drainage and b) imbibition. The black lines denote single phase injection, and the blue lines denote co-injection. The red dashed line indicates a spectral slope of -2, which was seen in other multiphase flow experiments \cite{spurin2022red, spurin2023pore}.}
	\label{FT_pressure}
\end{figure}

To explore timescales further, the pressure data from Figure \ref{pressure_data} was transformed into the frequency domain in Figure \ref{FT_pressure} using the methodology described in \citet{spurin2022red}. In Figure \ref{FT_pressure}, the red dashed line denotes a slope of 1/$f^2$ i.e. red noise. This dashed line fits the imbibition data for both experiments, with the single phase imbibition power spectra having a larger amplitude, but the same slope as the co-injection imbibition power spectra. The slope showing red noise fits the longer timescales (lower frequencies) during drainage, but the power spectra flattens out at the shorter timescales (higher frequencies) to white noise. 

Figure \ref{FT_pressure} shows that the co-injection drainage power spectra has a larger amplitude than the single phase injection drainage power spectra. This is reversed during imbibition; with the single phase imbibition power spectra having a larger amplitude than co-injection imbibition power spectra. Both imbibition power spectra have a larger amplitude than the drainage power spectra. While there is a greater difference in the pressure difference across the core, and the saturation along the core, for the drainage sequence for both experiments, the power spectra have a more similar amplitude when compared to the imbibition sequence for both experiments where the saturation across the core are similar. The end effects evolve over longer timescales and, as shown in Figure \ref{FT_pressure}, the longer timescales have a larger amplitude in the power spectra, and thus are influencing the pressure data and relative permeability calculations more. This highlights the importance of understanding the manifestation of end effects in the pressure data and associated relative permeability values. 

\subsection{Impact of experimental methodology on porosity-saturation relationships}

\begin{figure}[h]
	\begin{center}
		\centering
		\includegraphics*[angle=0, width=0.95\linewidth]{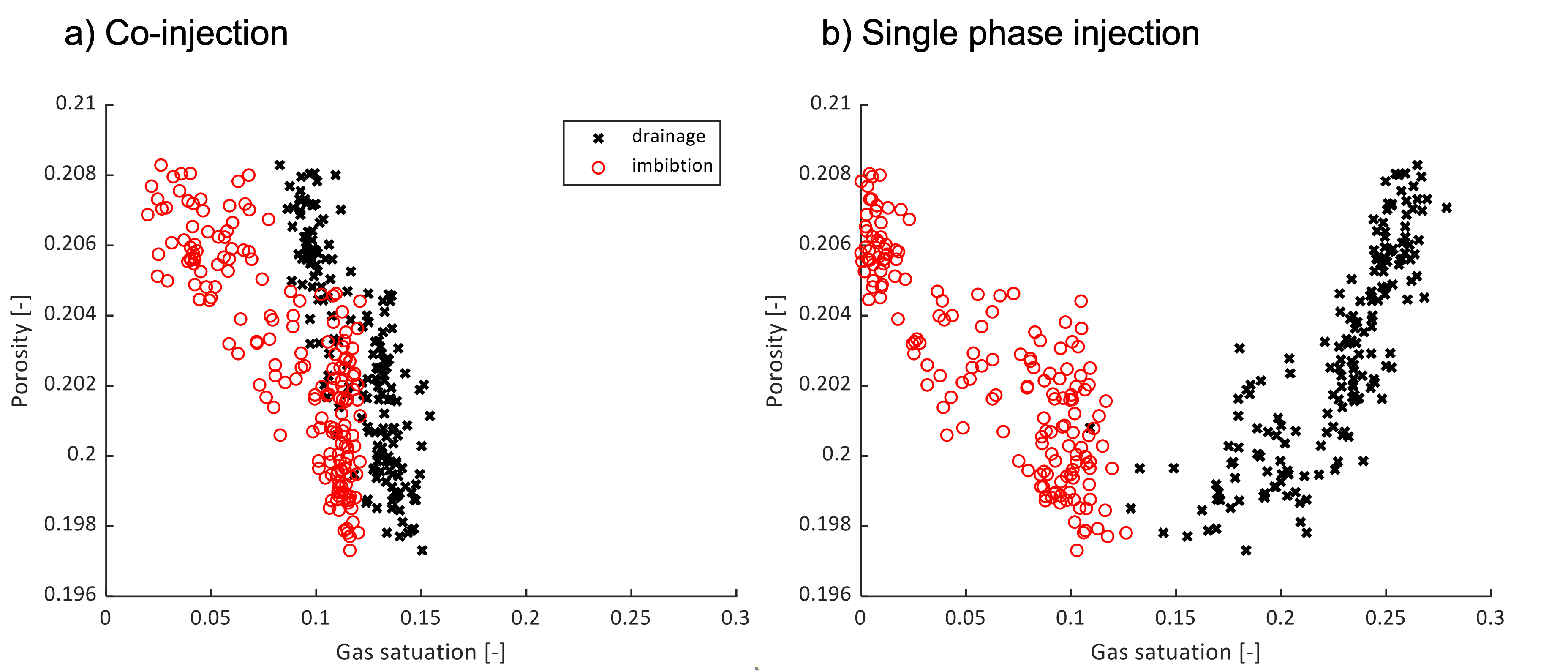}
	\end{center}
	\caption{The relationship between saturation and porosity for each slice in the core. The black crosses denote drainage, and the red circles denote imbibition for a) the co-injection experiment and b) the single phase injection experiment.}
	\label{sat_poro_n2}
\end{figure}

A strong saturation gradient was evident for the gas experiments near the inlet. It was hypothesized that the region affected could be quantified by measuring the change in saturation with time. To explore the potential impact of this effect on observations, the gas saturation is plotted against porosity in Figure \ref{sat_poro_n2}. As gas is the non-wetting phase, it would be expected to occupy the largest pores \cite{bluntbook, bultreys2018validation}. Thus, gas saturation would be expected to be higher in more porous regions i.e. a positive correction between gas saturation and porosity. This is only seen for single phase drainage. For all other cases, there is a negative correlation between gas saturation and porosity. For this sample (Indiana core 1 in Figure \ref{porosity}), sample porosity was higher closer to the inlet, so porosity was correlated with distance along core. This means that dissolution effect opposes the relationship expected from the porosity. This seemingly overrides the expected relationship between porosity and gas saturation for both drainage and imbibition during the co-injection experiment in Figure \ref{sat_poro_n2} a), even when not evident in the trapping efficiency in Figure \ref{trapping_curve}.

This suggests that the boundary conditions influence the relationship between gas saturation and porosity extends across the whole core, even at the outlet where there is no strong saturation gradient. This suggests that analyses that crop the end effects out, will still be influenced by the end effects, highlighting the need to understand the impact of experimental methodology on trapping in order to successfully apply core scale experiments to subsurface reservoir models. The impact of experimental methodology must be carefully considered if one is using porosity-saturation relationships to deduce the wettability of the core. Furthermore, porosity-saturation relationships may not be a reliable measurement.

\subsection{The role of heterogeneity on trapping efficiency}

\begin{figure}[h]
	\begin{center}
		\centering
		\includegraphics*[angle=0, width=0.9\linewidth]{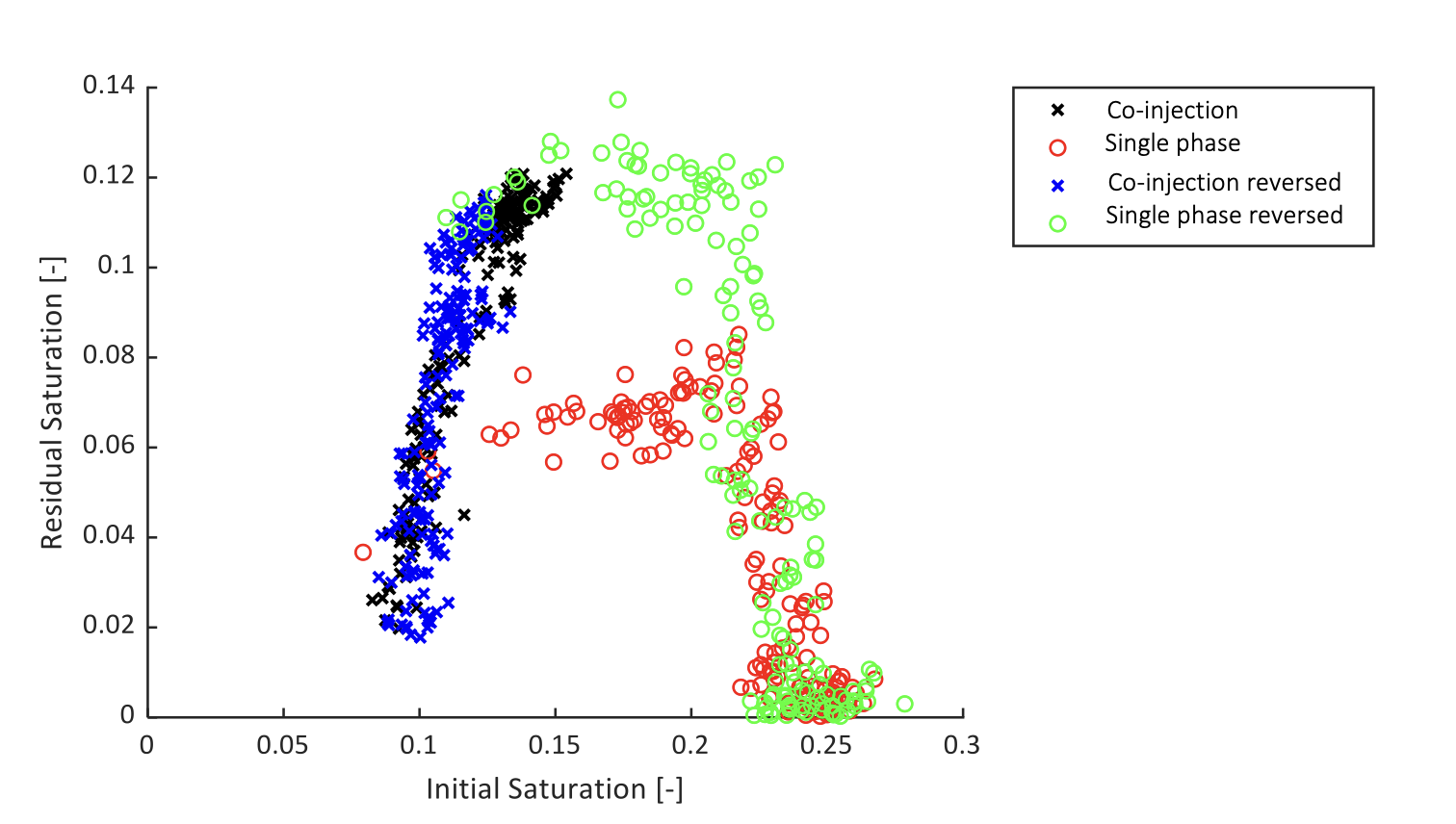}
	\end{center}
	\caption{The relationship between between initial gas saturation (post drainage) and residual gas saturation (post imbibition) for the co-injection experiment (black and blue crosses, the latter for the reversed core orientation) and single phase injection experiment (red and green circles, the latter for the reversed core orientation).}
	\label{trapping_curve_rev}
\end{figure}

The core for the gas experiments was reversed to assess the impact of the orientation of heterogeneity on flow properties for the different injection techniques. There were variations in saturation across the core (which were expected). The important difference is in the trapping curves in Figure \ref{trapping_curve_rev}. For the co-injection experiments, the trapping curve remains the same, regardless of core orientation. However, for the single phase experiments, the reversed core orientation leads to a different trapping curve, caused by the dissolution effects becoming more influential. This is likely due to the now slightly decreasing porosity with distance from the inlet; the porosity gradient is working in the same direction as the end effect, whereas for the initial orientation, the different effects were opposing each other. Thus, the role of heterogeneity was dependent on the injection technique. 

The role of heterogeneity with different injection technique may differ for the oil experiments. Likely, the orientation will have a smaller impact because there was less difference between the different injection techniques. This was not tested as it is difficult to remove oil from the sample. However, it is likely that small-scale heterogeneities are more important for gas experiments, compared to traditional oil experiments \cite{Spurin2025commentary}, highlighting the need to design different experiments for carbon storage projects. 

\section{Conclusions}

Significant time and effort has, historically, been invested in measuring relative permeability and trapping efficiency in reservoir rocks under representative conditions. These measurements have traditionally employed steady-state co-injection methodologies specifically designed to allow for a constant saturation profile with time, and to minimize capillary end effects. This study demonstrates that injection technique significantly influences both pressure response and trapping efficiency—factors that must be properly accounted for when upscaling laboratory results to field applications. The findings emphasize that experimental design fundamentally shapes observations, a critical consideration when translating laboratory data to field-scale implementations.

The orientation of heterogeneities exerted significant influence on trapping efficiency during single-phase injection experiments, while showing less pronounced effects in co-injection experiments. This indicates that heterogeneity classification and associated uncertainties become considerably more significant when employing single-phase injection protocols. Furthermore, failure to account for flow complexity and the manifestation of boundary conditions in trapping relationships and pressure measurements will inevitably lead to inaccurate reservoir characterization. The injection methodology produced more significant effects in gas experiments compared to oil experiments, further emphasizing why CO$_2$ storage projects require modeling frameworks distinct from those conventionally applied to oil reservoirs.

The co-injection of water with gas demonstrably enhances trapping efficiency, even when excluding boundary effects from analysis. However, this approach reduces pore volume utilization, resulting in greater lateral plume migration despite increased residual trapping. Consequently, a comprehensive assessment of the trade-off between plume spread and residual trapping is essential to determine whether co-injection or water-alternating-gas (WAG) strategies would prove beneficial at field scale. Overall, these results underscore the critical importance of considering experimental design influences when interpreting laboratory observations and extrapolating findings to field-scale implementation.

\bibliography{refs}

\end{document}